\documentclass[prbl,twocolumn,preprintnumbers]{revtex4}

\usepackage{amsmath}
\usepackage{amssymb}
\usepackage{bbold}
\usepackage{mathrsfs}
\usepackage{graphicx, psfrag}
\usepackage[centerlast]{caption}
\usepackage{float}
\usepackage{subfig}
\usepackage{tikz}
\usepackage{pgfplots}
\usepackage[colorlinks=true, citecolor=blue, urlcolor = blue, linkcolor= red, bookmarks=true]{hyperref}
\usepackage{epstopdf}
\usepackage{afterpage}
\begin{document}
\newcommand{\ts}{\textsuperscript}
\def \beq{\begin{equation}}
\def \eeq{\end{equation}}
\def \bse{\begin{subequations}}
\def \ese{\end{subequations}}
\def \bea{\begin{eqnarray}}
\def \eea{\end{eqnarray}}
\def \bs{\boldsymbol}
\def \bb{\bibitem}
\def \nn{\nonumber}
\title{\textbf{THz Photodetector using sideband-modulated transport through surface states of a 3D
Topological Insulator }}
\author{Puja Mondal$^{1}$, Sankalpa Ghosh$^{1}$ and Manish Sharma$^{2}$}
\affiliation{$^{1}$Department of Physics, Indian Institute of Technology Delhi, New Delhi-110016, India}
\affiliation{$^{2}$ Pitney Bowes Software India Pvt. Ltd., Noida, U.P-201303, India}
\begin{abstract}
The transport properties of the surface charge carriers of a three dimensional topological insulator under a terahertz (THz) field along with a resonant double barrier structure is theoretically analyzed  within the framework of Floquet theory to explore the possibility of using such a device for photodetection purpose.  We show that due to the contribution of elastic and inelastic scattering processes in the resulting transmission sidebands are formed in the conductance spectrum in somewhat similar way as in an optical cavity and this information can be used to detect the frequency of an unknown THz radiation.  The dependence of the conductance on the bias voltage, the effect of THz radiation on resonances and the influence of zero energy points on the transmission spectrum are also discussed. 
\end{abstract}
\maketitle
\section{Introduction}
Three-dimensional topological insulators (3DTIs) have generated a lot of interest in the field of condensed matter physics and material science due to their unique surface properties \cite{Hasan-a, Qi-a}. The low energy quasi-particles on the surface of such 3DTIs are massless Dirac fermions with chiral spin texture \cite{Zhang-a, Hsieh-a, Xia-a, Chen, Hsieh-b}.  
 The relevant energy scale in the case of 3DTIs lie in the terahertz (THz) region and there have been already interesting reports on the THz radiation from the bulk \cite{Bristow}.  Some of the developments in the direction of THz related studies are helicity-controlled photocurrents \cite{McIver,Junck,Hamh1}, THz quantum Hall effect of Dirac fermions \cite{Shuvaev}, giant photocurrent at cyclotron resonance in a THz radiation \cite{Olbrich2, Dantscher}, magneto-oscillations of THz radiation-induced photocurrent \cite{Zoth}, anisotropy photogalvanic effects by the THz electric field \cite{Olbrich1}, and tunable photogalvanic effect via proximity interaction with illumination of THz radiation \citep{Semenov1,Semenov2}. 
 
 Since surface states are expected to have strong absorbance and high signal-to-noise ratio (SNR)  in the THz regime \cite{Xiao}, they can be used as a THz radiation detector \cite{Luo}. However,  relatively less application-oriented work has taken place in this direction.
 THz radiation with photon energy lower than the bulk band gap affects only the surface states without creating any bulk excitations. 
On the other hand, there has been interesting progress in the study of transport properties of the two-dimensional material under the influence of periodic time-dependent electromagnetic (E.M.) field   \cite{Shirley,Lindner-Refael}. Treating the interaction of such surface states of 3DTI with an incident THz radiation in the framework of Floquet theory, we propose a tunable optoelectronic model that can be used as a photodetector for THz radiation.

We arrange the manuscript as follows. In section \ref{dev-app}, we introduce the structure of the device and how its properties can be used for the purpose of photodetection in THz regime in detail.  We also discuss in detail various properties of the transmission spectrum and the corresponding conductance in presence of zero energy points which can be applied for making a finite frequency noise detector.  
 In section \ref{method}, we discuss the methodology to obtain the conductance of the surface electrons in presence of the gated double barrier structure and THz field within the framework of Floquet theory. Conclusions and discussion are presented in section \ref{conclu}.

\begin{figure*}
\subfloat{\includegraphics[width=2\columnwidth,height=1.2
\columnwidth]{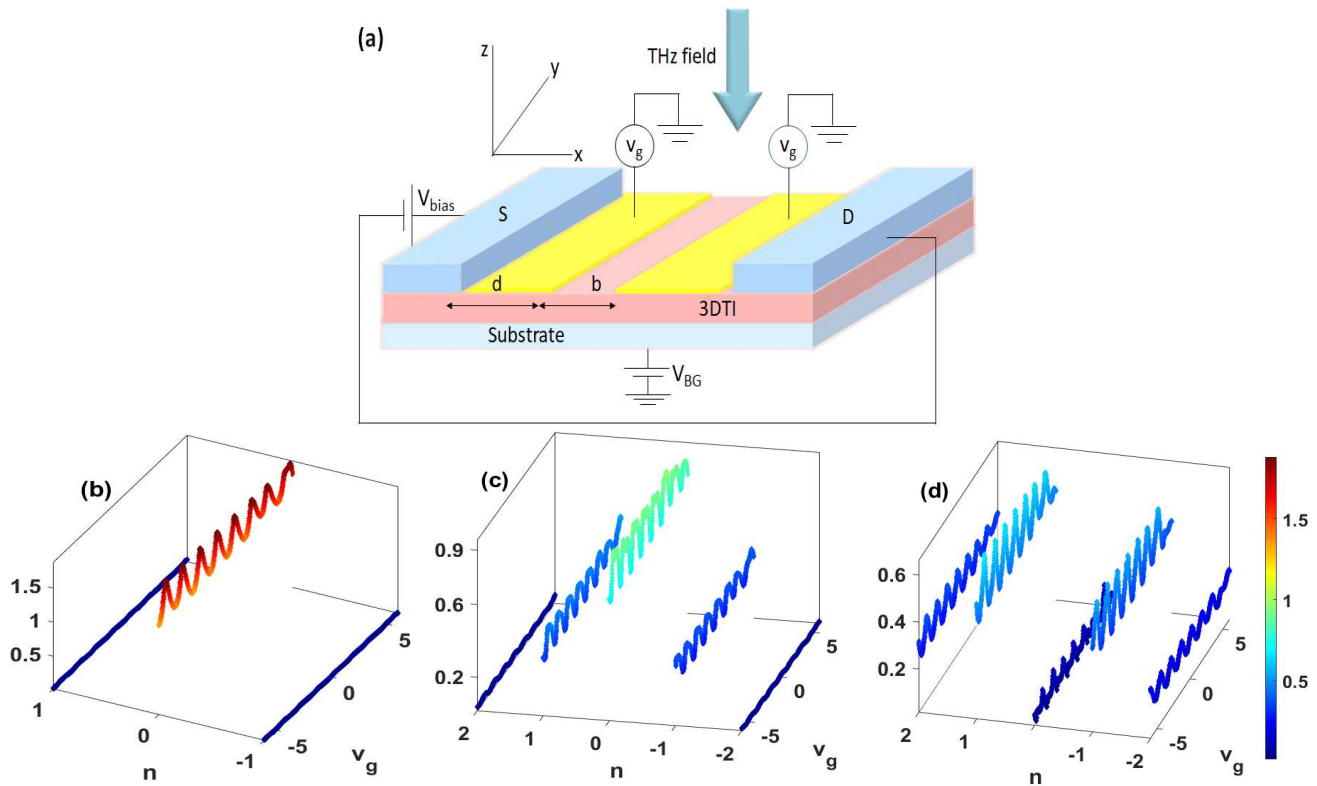} }
\caption{ (a) Schematic of the system. Blue regions depict the left and right ohmic contacts. A bias voltage is applied between the left and right leads. Yellow regions correspond to the top gated areas producing a scalar potential on the surface. $d$ and $b$ are the barrier width and separation. Back gate is used to tune the fermi energy of the bare system. (b), (c) and (d) are central ($ n=0 $) and side bands ($ n\neq 0 $) conductance vs effective barrier strength ($ v_g $) for the parameters given in Fig. \ref{fig:condvspotential_bias_vary} (a). }
\label{fig:schematic}
\end{figure*}
\section{ device and its application} \label{dev-app}
The device we propose is a gated structure on the surface of a 3DTI in the form of a double barrier potential \cite{Mondal} sketched in Fig. \ref{fig:schematic} (a).   Our model contains three regions: one central region tuned by the back gate voltage, two gated areas (yellow areas) tuned by both back and top gate and ohmic contacts (source and drain). The detailed discussion on the experimental realization of the proposed model is given in section \ref{method}.   Here, the double barrier serves the purpose of an electronic analogue of the Fabry-P\'erot cavity where the p-n junctions play the role of mirrors. 
But with an added advantage over the optical Fabry P\'erot cavity, namely that one can tune the fermi wavelength of electrons by changing the gate voltage. 

Such an electronic Fabry-P\'erot cavity produces resonances in the transmission due to the reflection and transmission of the electronic wave function at the junctions. These Fabry-P\'erot resonances are tunable by changing the cavity length and the fermi wavelength of the electron \cite{Sharma, Masir, Rickhaus, Varlet, MTAllen}.
 The Fabry-P\'erot resonances have been a useful tool to investigate different kinds of properties of the system such as evidence of broken chirality in Fabry-P\'erot interferences \cite{Varlet1}, a $\pi$ phase shift in the interference fringes in a magnetic field \cite{Shytov, Young}.  We consider cavity length of the order of 100 nm which remains constant throughout our analysis where cavity length is defined as the separation between two barriers. The barrier width is of the order of 50 nm.

We then allow a beam of THz radiation to interact with the electronic Fabry-P\'erot cavity on the surface of a 3DTI. The THz radiation stimulates inelastic scattering processes in the system in which quanta of photons are absorbed or emitted by the electrons, as evidenced experimentally by Dayem and Martin \cite{Dayem} and theoretically analyzed by Tien and Gordon \cite{Tien}. 
This creates new transmission channels corresponding to energies $ E+n\hbar\omega $ in the system as shown in the Fig. \ref{fig:schematic}. In Fig. \ref{fig:schematic}, $n=0$ transmission channel corresponds to the case when the energy of the incident and transmitted electrons remains unchanged. For $n>0$, electrons absorb $  n$ quanta of photon and for $n<0$, electrons emit $ n $ quanta of photon while passing through the system.  Thus, there is a certain probability that the electron with initial energy E will be scattered to the final energy states with energy $E+n\hbar\omega$, where $E+n\hbar\omega$  is the energy of the Floquet subbands of the system.

As demonstrated in Fig. \ref{fig:schematic}, the central band transmission probability gets reduced as the drive strength of the field increases (where the drive strength $Z=\frac{V_t}{\hbar\omega}$), while the Floquet sideband transmission probability gets amplified. Here, one needs to mention that in the process of increasing drive strength, the frequency of the THz radiation is kept constant.
The Floquet sideband formation in the transmission/conductance (Fig. \ref{fig:schematic} )
can be understood as an electronic analogue to sideband formation in 
a prototype cavity optoelectronic system where the pump laser frequency is modulated by the mechanical frequency \cite{Schliesser, Kippenberg, Teufel, Schliesser1, Aspelmeyer}. In this case, the Fabry-P\'erot resonance energy (the energy at which transmission becomes unity) gets modulated by the external THz radiation and forms $  n$ number of sidebands where $ n $ depends on the drive strength.
\begin{figure}
\subfloat{\includegraphics[width=1\columnwidth,height=1.2
\columnwidth]{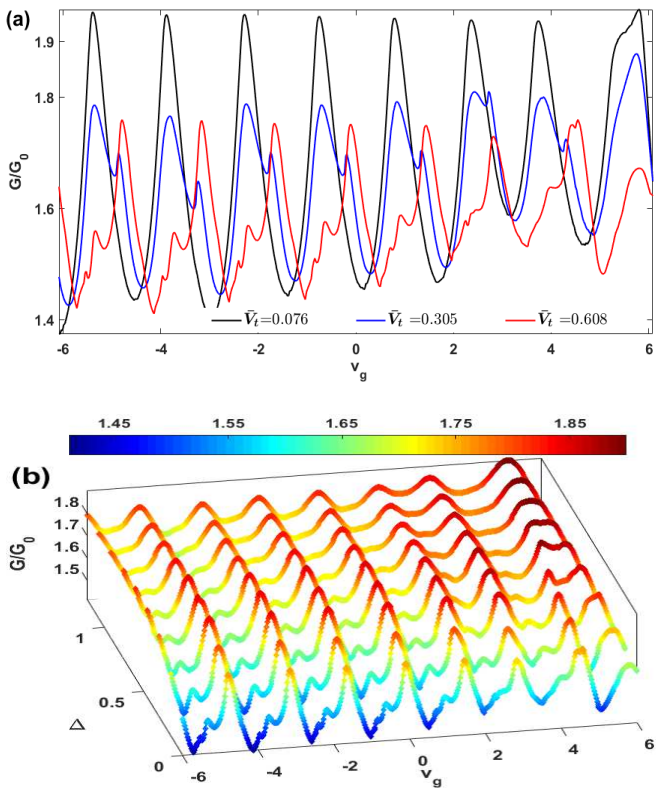} }
\caption{ (a)  Conductance versus effective strength of scalar potential for a fixed frequency $\bar{\omega}=$0.305 and different values of amplitude of the time dependent potential at zero bias. (b) Conductance vs effective barrier strength ($v_g$) for the increasing bias between the left and right contacts ($ \Delta$). We consider $\bar{V}_t$ =0.608 and $\bar{\omega}$=0.305. The corresponding value of Z is 2.  The energy is measured in units of $ \hbar v_{F}/d $= 6.57 meV. Frequency is measured in units of $ v_{F}/d= 10^{13} $ Hz. We consider $\varepsilon=$ 7.610, $ b/d  $=2  and $ \bar{\mu}_{L}=\bar{\mu}_{R} $=6.088. }
\label{fig:condvspotential_bias_vary}
\end{figure}

An experimentally measurable quantity would be the total conductance which includes the contribution of all bands (central band and sidebands) (shown in Fig. \ref{fig:condvspotential_bias_vary}). The total conductance is expected to exhibit an oscillating behavior as a function of effective barrier strength. The pattern of the oscillations in the conductance would get modified as Z increases.    The black curve in Fig. \ref{fig:condvspotential_bias_vary} corresponds to weak drive strength which shows oscillations as a function of effective barrier strength ($v_g$). These oscillations occur due to the presence of Fabry P\'erot cavity and oscillations take the Breit-Wigner distribution form \cite{Levitov, Shevtsov}. Phenomenologically, one can write the conductance as $G(v_g) \propto q G_0 \Gamma^{2}/\bigg((v_g-v_g^{r})^{2}+\Gamma^{2} \bigg )$ where q is a pre-factor (constant), $v_g^{r}$ is the effective barrier strength at which resonance occurs and $\Gamma$ is the width at half maxima.  An interesting effect arises when we increase drive strength: extra resonance peaks appear in the conductance and alter the Breit-Wigner distribution form of the conductance (as appeared in Fig. \ref{fig:condvspotential_bias_vary} (a) for $ \bar{V}_t $=0.608).   

The amplitude of extra resonance peaks increases as drive strength becomes stronger and  modifies the Breit-Wigner distribution form of the conductance. Thus, at higher Z, one can write the conductance as 
\beq G(v_g) \propto \sum_n \frac{q(n,z) G_0 (\Gamma(n,z))^{2}}{\bigg((v_g-v_g^{r}(n,z))^{2}+(\Gamma(n,z))^{2} \bigg )} \eeq
This means we can use this model as a highly tunable THz detector. The extra resonance peaks appear in the conductance due to inelastic scattering processes which are nothing but the photon-assisted scattering (PAS) processes \cite{Li, Trauzettel, Zeb, Biswas, Zambrano}. In PAS process, discrete photon energy ($\hbar \omega  $) can be detected when an electron absorbs or emits $n$ photons and scatters to the final state where $\varepsilon_{f}-\varepsilon_{i}=n\hbar \omega  $.   The extra resonance peaks emerge at values of barrier strength different from the main resonance peaks (Fabry P\'erot resonance peaks without THz radiation), which solely depend on the frequency of the THz radiation.  By tuning the gate voltage, one can measure the frequency of an unknown THz radiation.  The result obtained in Fig. \ref{fig:condvspotential_bias_vary} (a) is for zero bias voltage.

The proposed device can be operated in two possible modes. First, to adjust drive strength, we change the bias applied to the right and left leads \cite{MTAllen}. Now, as we increase the bias between  two leads ($ \Delta=\bar{V}_{L} -\bar{V}_{R} $)  at a fixed $  Z$=2 (as shown in Fig. \ref{fig:condvspotential_bias_vary} (b), the amplitude of the main resonance peaks (peaks without THz radiation) gets reduced which corresponds to elastic scattering process. The remaining resonances are due to inelastic scattering process  where $n$ number quanta of photon contribute.   This can be understood from the central and side-bands conductance which is plotted in Fig. \ref{fig:tran_sideband_Vt4meV_omega_vary} (c) for $ \Delta$ =1.2164. At $ \Delta$ =1.2164, the central band conductance is very small as compared to side conductance. In this case, conductance peaks of $T_0$, $T_1$ and $T_{\pm 2}$ oscillate in phase. Whereas, conductance peaks of $T_{-1}$ oscillate out of phase with the rest. At lower bias voltage, conductance peaks of $T_1$ and $T_2$ oscillate in phase while oscillate out of phase with $T_{-1}$. But, in this case, there is a phase difference between oscillation peaks of $T_{-2}$ with $T_{1,2}$ and  $T_{-1}$. Thus, at higher bias voltage, in phase oscillation of conductance peaks of $T_0$, $T_1$ and $T_{\pm 2}$ (where conductance due to $T_0$ is insignificant) give prominent conductance peaks at $T_1$ while zero conductance peaks at $T_{-1}$ and$T_{0}$.    
Therefore, we can tune $  Z$, $ \bar{V}_{L} $ and $ \bar{V}_{R} $ to separate out the elastic and non-elastic contributions of the scattering inside the device. 
 \begin{figure}
\subfloat{\includegraphics[width=1\columnwidth,height=1.2
\columnwidth]{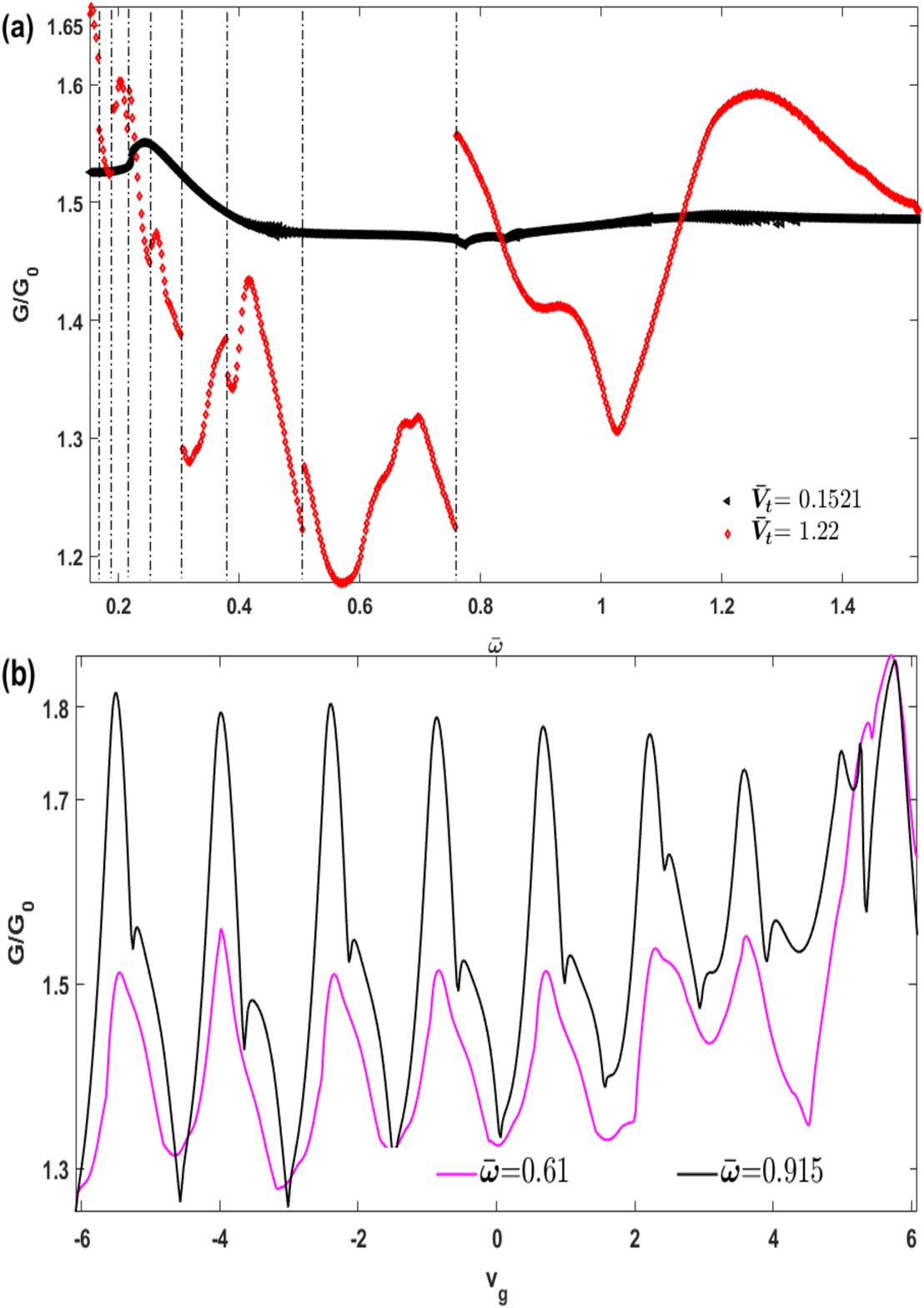} }
\caption{ (a) Conductance vs frequency of the time dependent potential for a fixed value of effective barrier strength ($v_g$=1.5205) and the amplitudes of the time dependent potential are $\bar{V}_{t}$= 0.1521 and $\bar{V}_{t}$= 1.22. The dotted lines are showing discontinuity points where the conductance are not defined. (b) Conductance vs $v_g$ for two different values of frequency $\bar{\omega}$=0.601 and 0.915 for a fixed $\bar{V}_{t}$= 0.608. The other parameters are same as in Fig. \ref{fig:condvspotential_bias_vary} (a).     }
\label{fig:tranvsomega}
\end{figure}
  
\begin{figure}
\subfloat{\includegraphics[width=1\columnwidth,height=1.2
\columnwidth]{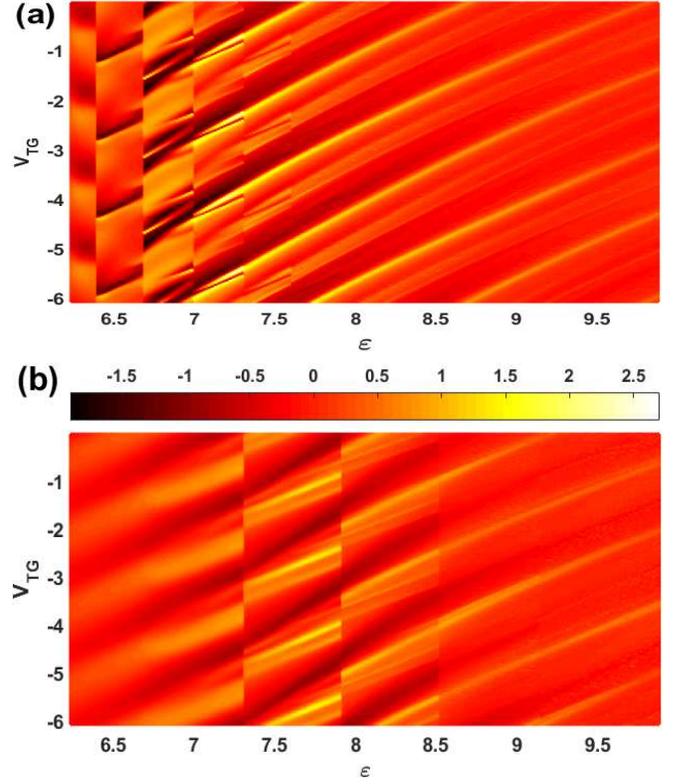} }
\caption{ (a) Differential conductance ($dG/dV_{TG}$) vs energy ($ \varepsilon $) and top gate voltage($ V_{TG} $) for $\bar{V}_{t}$=0.608 and $\bar{\omega}$=0.305 and (b) for $\bar{V}_{t}$=1.22 and $\bar{\omega}$=0.610 at drive strength $Z$=2.  The value of $\varepsilon$ can be tuned using the back gate. Differential conductance shows discontinuity whose position changes as the frequency changes. The unit of the energy is  $ \hbar v_{F}/d $= 6.57 meV and the unit of  the frequency is  $ v_{F}/d= 10^{13} $ Hz. We consider  $ b/d  $=2  and $ \bar{\mu}_{L}=\bar{\mu}_{R} $=6.088.    }
\label{fig:differential_cond}
\end{figure}

\begin{figure*}
\subfloat{\includegraphics[width=2\columnwidth,height=1.2
\columnwidth]{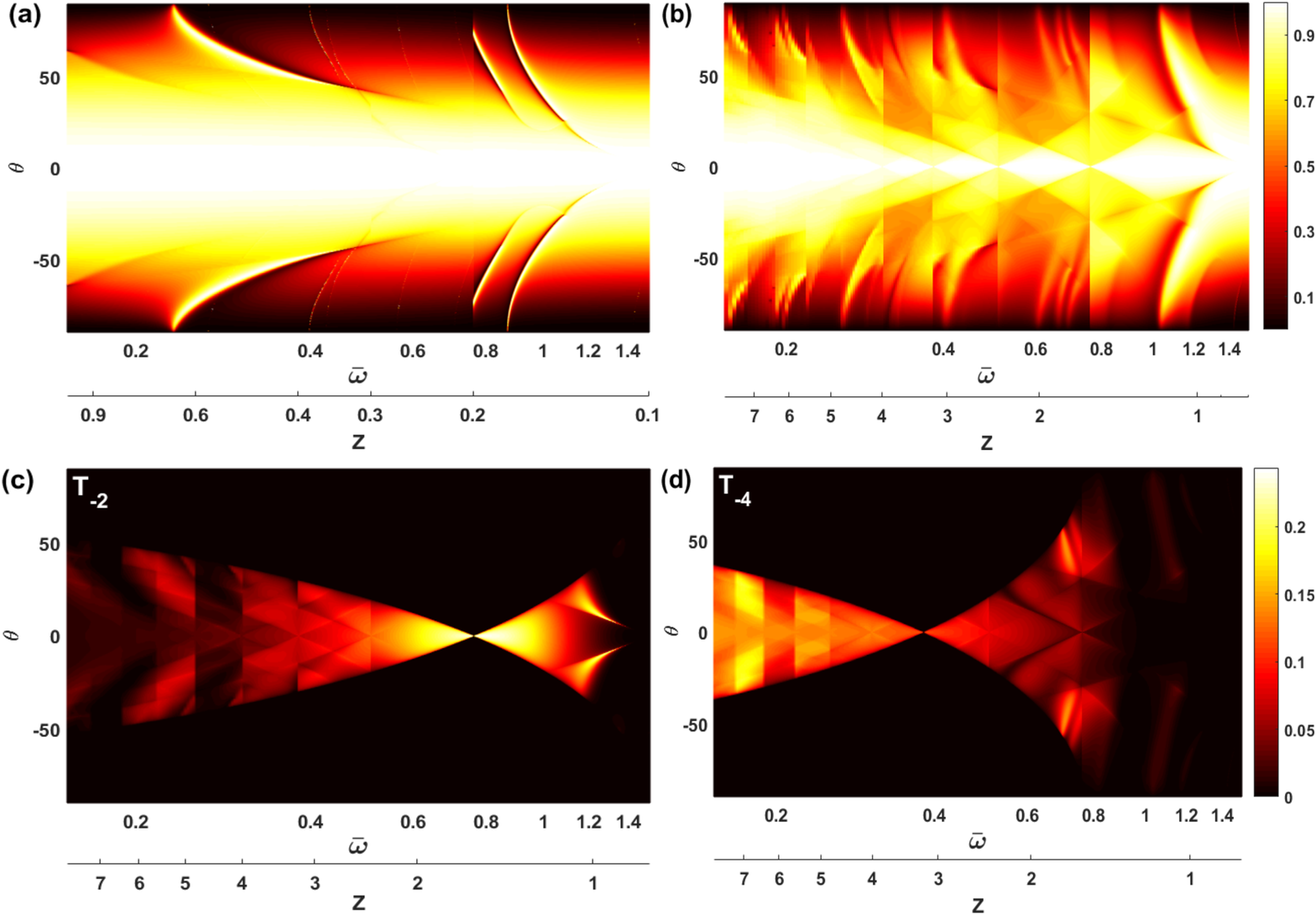} }
\caption{(a) and (b) are the transmission plot versus angle of incidence and frequency corresponding to Fig. \ref{fig:tranvsomega} (a) for $\bar{V}_{t}$= 0.1521 and $\bar{V}_{t}$= 1.22.  (c) and (d) show the transmission of the side bands corresponding to Fig. b ($\bar{V}_{t}$=1.22) for $n$=-2 and $n$=-4. The other Parameters used in this figure are same as Fig. \ref{fig:condvspotential_bias_vary} (a). }
\label{fig:tranvsomega_side_bands}
\end{figure*}

 Thus, the elastic scattering process gets highly suppressed when the drive strength becomes strong ($ Z > 1 $). The prominent contribution to the conductance comes from the inelastic scattering processes where the Floquet side-bands are involved in the scattering processes. In this way by combining effect of the gate and bias voltage, one can experimentally measure the dc response of the surface electrons solely due to photon assisted charge carriers. Therefore, this proposed model can be used as highly tunable photodetector in the THz regime.   
For a signal with unknown frequency, one can detect the frequency from the sidebands conductance as sideband conductance peaks are separated from the dc conductance peaks. 

 The second mode of operation of the proposed device is as follows. In order to vary Z, we can keep $\bar{V}_t$ constant while changing $\bar{\omega}$. Usually, performing conductance measurements while sweeping over a frequency range is a standard measurement to set up \cite{Luo}.  To show the effect of Z(or $\bar{\omega}$) on the conductance, we demonstrate transmission and conductance vs Z in Fig. \ref{fig:tranvsomega} (a) for a fixed $ v_g $ and corresponding transmission is plotted in Fig. \ref{fig:tranvsomega_side_bands}. At smaller $\bar{V}_{t}$, the conductance shows smooth behavior. But as $\bar{V}_{t}$ increases, conductance shows discontinuity at some particular value of $\bar{\omega}$. At these points, the conductance is not well defined. The discontinuity points increase with increasing $\bar{V}_{t}$ and at $\bar{V}_{t}$=1.22, there are several such discontinuity points in the conductance (shown in Fig. \ref{fig:tranvsomega} (a)).  The discontinuity in the conductance arises when 
\beq
\varepsilon-\bar{\mu}_{R}+n\bar{E}_t=0. 
\label{eq:zero_energy}
\eeq
This happens when zero energy states emerge in the system. 

For zero energy states, the corresponding solution of the wavefunction is different (as shown in the appendix \ref{zero_energy}). The electronic states lie at the newly formed zero point energies where the density of states vanishes. There are such several zero energy points present whose position can be tuned either by changing the frequency or the other characteristic tunable parameters of the system.    One can compare this phenomenon with the emergence of zero energy modes in a spatially periodic potential \cite{Brey, Park} and also in a time-periodic potential \cite{Savelev1, Savelev2}. The emergence of these new zero energy points is still there even at finite bias between the leads. 

The zero energy states appear at the Floquet bands lie at energies lower than the central band ( {i.e} for $n$=-1,-2,-3 {\it etc}). But for a distinct Floquet band, zero energy states appear at a specific value of $\bar{\omega}$. 
As an example, for $n=-1$, $\varepsilon-\bar{\mu}_{R}+n\bar{E}_t$ becomes zero at  $\bar{\omega}$=1.525 as we consider $\varepsilon-\bar{\mu}_{R}$=1.525. Similarly, $\varepsilon-\bar{\mu}_{R}+n\bar{E}_t$ becomes zero at  $\bar{\omega}$=0.7625, 0.5083,  0.3812,  0.3050, 0.2542 for $n=-2,-3,-4, -5 -6 $. At these points, transmission of the corresponding Floquet side bands appears to be zero (presented in Fig. \ref{fig:tranvsomega_side_bands} (c) and (d)) as the propagating modes in the side band $n$ vanishes . 
  
\begin{figure*}[h!]
\subfloat{\includegraphics[width=2\columnwidth,height=0.70
\columnwidth]{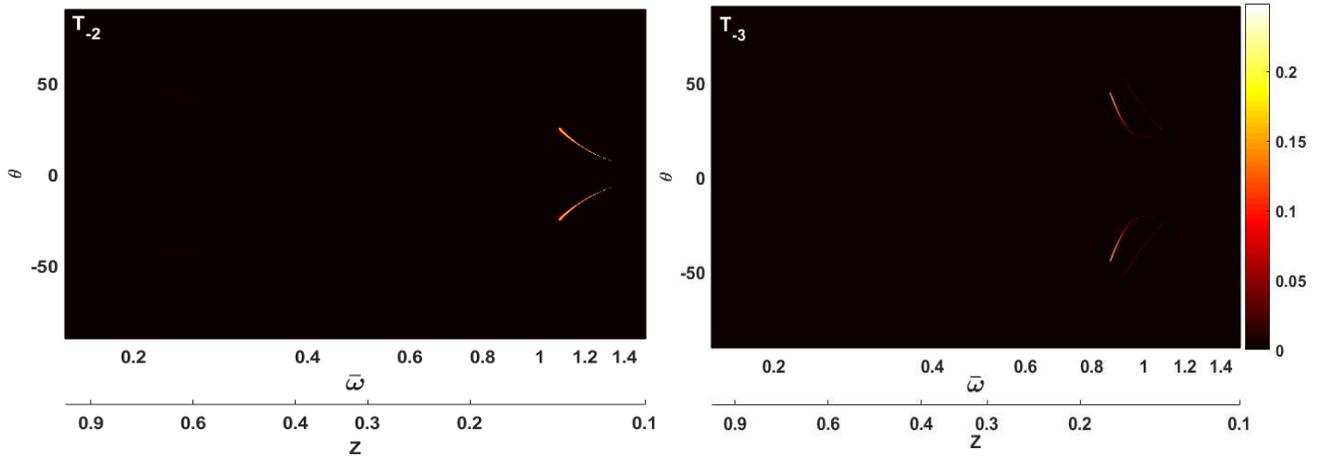} }
\caption{Transmission of $T_{-2}$ and $T_{-3}$ side bands corresponding to Fig. \ref{fig:tranvsomega_side_bands} (a).}
\label{fig:tran_sideband_Vt1meV_omega_vary}
\end{figure*}

While, at $\bar{V}_{t}$=0.1521, $Z<1$ which provides smaller or zero contribution of the Floquet side bands ($n\neq 0$) to the conductance (depicted in Fig. \ref{fig:tran_sideband_Vt1meV_omega_vary}).  Therefore, discontinuities in the conductance will be observable for strong drive strength. The discontinuity occurs when we measure conductance either varying frequency or the energy. To show the effect of varying energy (which is possible by tuning the back gate voltage), we calculate and plot differential conductance as a function of both back gate and top gate voltage in Fig. \ref{fig:differential_cond}. In Fig. \ref{fig:differential_cond} (a) and (b), we keep drive strength $ Z $ constant by changing both $\bar{\omega}$ and $\bar{V}_t$. We see that points of discontinuity shift as we change the frequency and the spacing between the discontinuities increases. Also, small resonance peaks are visible in both the cases. It is possible to experimentally measure $dG/dV_{TG}$ experimentally by modulating top gate voltage with an ac voltage of small magnitude \cite{Varlet}.  Such differential conductance measurements are quite routine in electrical transport studies of semiconductor devices, and it is well-known that they are very sensitive for measuring band structure properties and interface states.     
 One can use this feature of discontinuity in the conductance to detect the finite frequency noise in the system \cite{Deblock}. One limitation of using as a frequency noise detector would be that the frequency  be kept smaller than the bulk band gap of the system. 
 
\begin{figure*}
\subfloat{\includegraphics[width=2\columnwidth,height=0.70
\columnwidth]{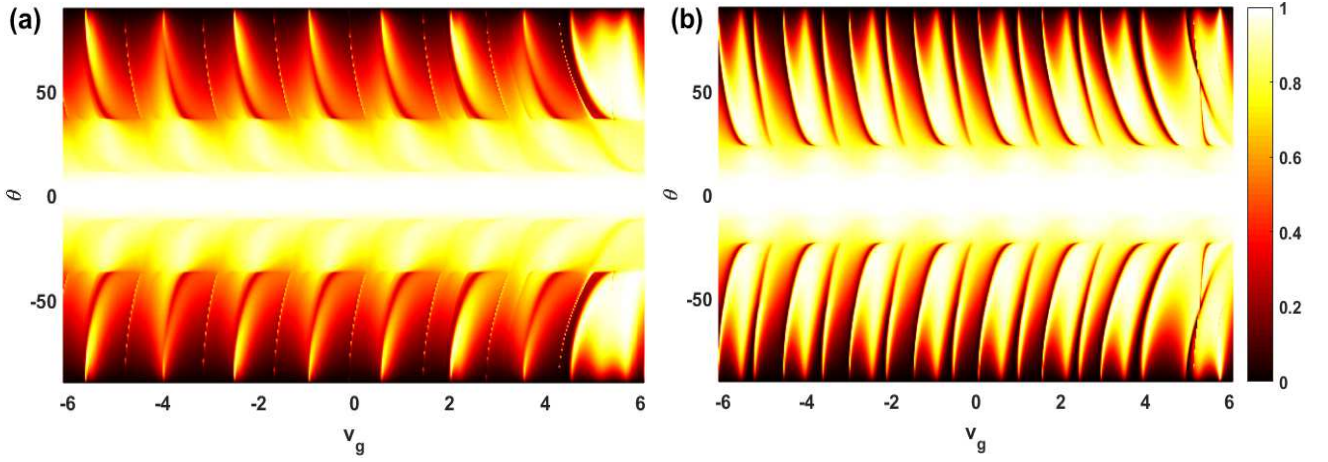} }
\caption{Transmission plot versus effective barrier strength of scalar potential for a fixed value of  $\bar{V}_t$= 0.608 and different values of frequency of time dependent potential (a) $\bar{\omega}$= 0.601 (b) $\bar{\omega}$= 0.915. The energy is measured in units of $ \hbar v_{F}/d $= 6.57 meV. Frequency is measured in units of $ v_{F}/d= 10^{13} $ Hz. We consider $\varepsilon=$ 7.610, $ b/d  $=2  and $ \bar{\mu}_{L}=\bar{\mu}_{R} $=6.088.}
\label{fig:tran_omega_vary}
\end{figure*}

\begin{figure*}
\subfloat{\includegraphics[width=2\columnwidth,height=1
\columnwidth]{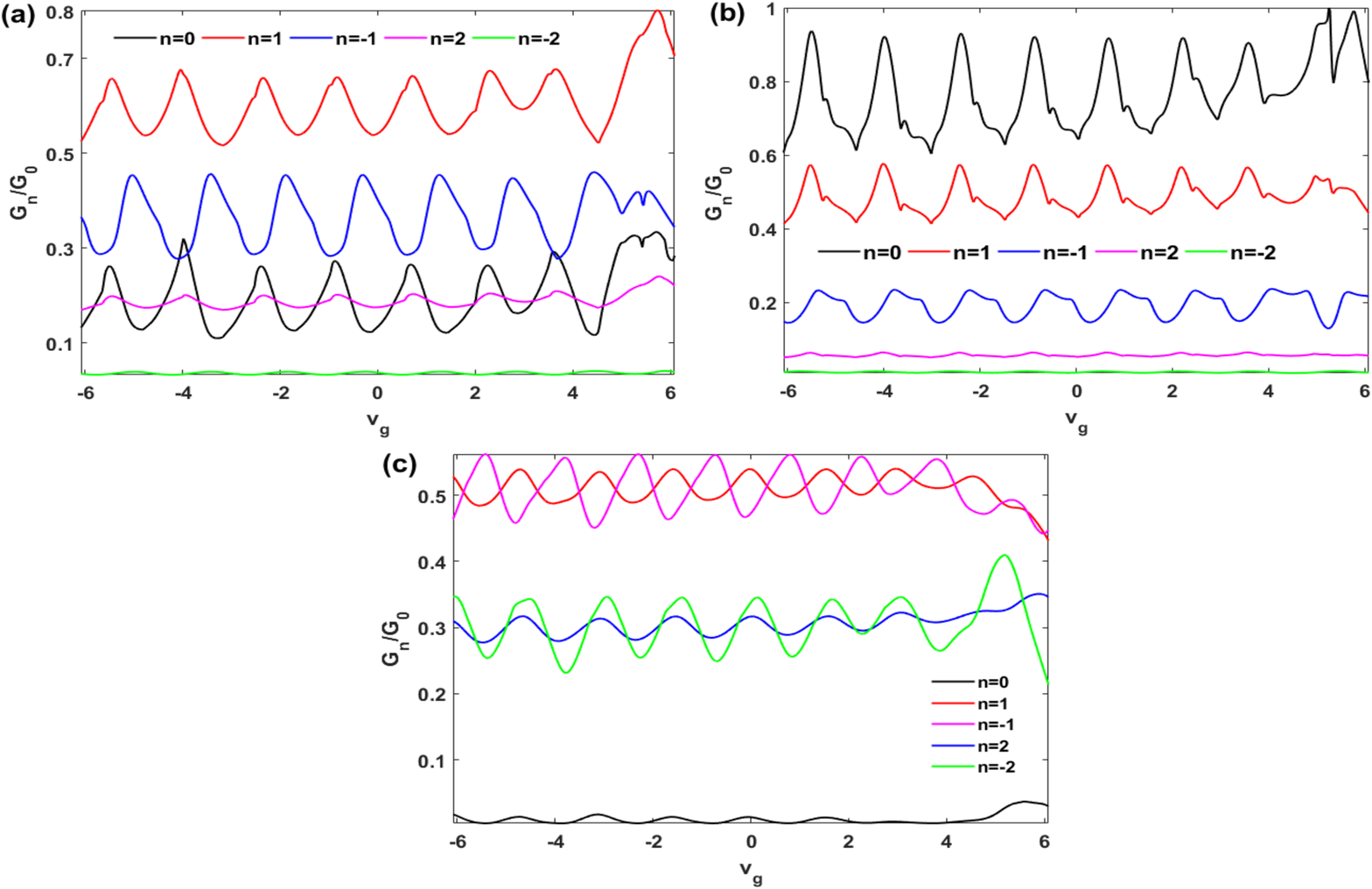} }
\caption{Conductance of the central band ($T_0$) and Floquet side bands ($T_{\pm 1, \pm 2}$) corresponding to (a) Fig. \ref{fig:tran_omega_vary} (a), (b) Fig. \ref{fig:tran_omega_vary} (b) and (c) Fig. \ref{fig:condvspotential_bias_vary} at $ \Delta$ =1.2164.}
\label{fig:tran_sideband_Vt4meV_omega_vary}
\end{figure*}

To study  the conductance vs $v_g$ for different drive strengths while keeping $\bar{V}_t$ constant, we vary the frequency of the THz radiation, and this is presented in Fig. \ref{fig:tranvsomega} (b). 
 The analogous transmission plot is shown in Fig. \ref{fig:tran_omega_vary}.  At $\bar{\omega}$=0.601, conductance peaks of $T_0$, $T_{1}$ and $T_{2}$ oscillate in phase (as shown in Fig. \ref{fig:tran_sideband_Vt4meV_omega_vary} (a)) while they oscillate out of phase with $T_{-1}$. Comparing the conductance peaks at $\bar{\omega}$=0.305 and 0.601, the peaks of $T_{0,1,2}$ at $\bar{\omega}$=0.305 (Fig. \ref{fig:schematic} (d)) oscillate out of phase with the peaks of $T_{0,1,2}$ at $\bar{\omega}$=0.601.  
 This results in shifting of the resonance position in the conductance. As in Fig. \ref{fig:tran_omega_vary} (c) for $\bar{\omega}$=0.601, the higher amplitude peaks come from $T_0$, $T_{1}$ and $T_{2}$ and smaller amplitude peaks come from $T_{-1}$, the position of which shifts as one decreases $\bar{\omega}$ from 0.601 to 0.305. As we further increase $\bar{\omega}$ small amplitude peaks in the conductance disappear.  This is true because the resonance condition changes as $\bar{\omega}$ varies.            
\section{ methodology}\label{method}
Now that we have explained the basic structure of the proposed device and its application in detail, we shall now briefly mention the methodology that was used to do the calculation to arrive at the results. Since double barrier configuration varies along the $ x $ direction, particle motion is free in the $  y$-direction.  The THz field produces a time-dependent potential with frequency $\omega$ and amplitude $V_{t}$ along the same direction as the scalar potential barrier. Also, we consider a metallic source and drain contact whose effect is considered as a change in the chemical potential at the left and right lead as shown in the Fig.   
 \ref{fig:schematic} (a). The THz field does not affect the leads region of the system for which one can consider the thick material having lower skin depth. While one can choose a thin material having higher skin depth for the gated region such as 1 nm thick gold plate. 
 
  We provide the dimensions of the device which are experimentally realizable. The range of radiation explored in this paper is 1 THz to 10 THz which is feasible for experiments \cite{Tang}.  This involves 750 nm to 30 $\mu$m dimensions, which are easily realizable by optical lithography. In terms of the active layer, we propose using a 3DTI material such as BSTS obtained by exfoliation {\it etc} \cite{Lee, Seifert, Sacepe}. The 3DTI layer needs to be thick enough to achieve proper conduction \cite{Tang, Hamh2, Giorgianni}, and also thin enough to allow interaction with the incident THz radiation.  Once placed on an insulating substrate, a back gate voltage can be applied. The typical lateral dimensions of the electrodes and barrier widths required can be achieved by either optical lithography or, preferably, by e-beam lithography techniques. Care needs to be taken to protect the 3DTI layer since it is so thin, and also to insulate the top metal electrode appropriately (some issues are described in \cite{Lee}).  Finally, the top electrodes would be deposited in an e-beam evaporator using standard metal layers such as Ti 10 nm (adhesive layer)/ Au 100-200 nm (metal layer). The use of Au also shields partly the 3DTI material by absorbing the THz radiation such that it only interacts with the device in the active barrier region. To protect the interaction area near the barriers, an additional windowing layer may need to be deposited (not shown in Fig. \ref{fig:schematic} for the sake of simplicity).

The full potential landscape is of the form   
\beq
 V(x,t) =
\begin{cases}
V_{L} & \text{if } x <-d-b/2 \\
V_{0}+ V_{t}cos(\omega t)  & \text{if } -d-b/2 < x < -b/2\\
V_{t}cos(\omega t)  & \text{if } \mid x\mid < b/2\\
V_{0}+ V_{t}cos(\omega t)  & \text{if } b/2 < x < d+b/2\\
V_{R} & \text{if } x >d+b/2 \\
\end{cases}
\label{pot}
\eeq
where $V_{L(R)}$ is the chemical potential of left (right) lead and $ d $ is the barrier width, $  b$ is the separation between barriers, $V_{0}$ is the barrier height, $V_{t}$ is the amplitude of time dependent field and $\omega$ is the frequency of time dependent field. 

The Hamiltonian describing the surface states of three-dimensional topological insulators (TI's) is of the form
\beq
H=v_{F}(p_{y}\sigma_{x}-p_{x}\sigma_{y})+\frac{\lambda}{2}\sigma_{z}(\sigma_{+}^{2}+\sigma_{-}^{2})
\label{hamil-full}
\eeq 
where $v_{F}$ and $ \lambda $ are the fermi velocity and wrapping parameter of the surface states. $ \sigma_{x,y,z} $ are the Pauli matrices and $ \sigma_{\pm}=\sigma_x \pm \sigma_y $.  We consider $v_{F}$ =$ 5\times10^{5}$ for $Bi_{2}Se_{3}$. The Hamiltonian of the surface states acquires non-linear dependency on the momentum (second term in Eq. \ref{hamil-full}) as one moves away from the Dirac point \cite{Fu-a,Kuroda}. The energy contour is circular upto energy 200 meV for $Bi_{2}Se_{3}$ and 150 meV for $Bi_{2}Te_{3}$. So, near the vicinity of the Dirac point Hamiltonian becomes
\beq
H=v_{F}(p_{y}\sigma_{x}-p_{x}\sigma_{y})
\label{hamil-Dirac}
\eeq   
Thus, the time dependent Dirac equation in presence of the potential profile of  Eq. \ref{pot} becomes
\bea
i\hbar \frac{\partial \Psi(\bf{r},t)}{\partial t} &=& \hbar v_{F}\left(\begin{array}{cc}
 0 & -i\partial_{y}+\partial_{x}\\
  -i\partial_{y}-\partial_{x} & 0
\end{array} \right)\Psi(\bf{r},t)\nn\\
&+& V(x,t)\Psi(\bf{r},t)
\label{eq:sch-equ}
\eea 
Since the potential landscape is periodic in time, one can apply Floquet theorem to obtain eigenfunctions and eigenenergies of the system in presence of time dependent potential. We write the solution of Eq. \ref{eq:sch-equ} using Floquet theorem as \cite{Li}
\beq
\Psi(\bf{r},t)= e^{-iE_{F}t/\hbar}\phi(\bf{r},t)
\label{floquet}
\eeq 
where $E_{F}$ and $\phi(\bf{r},t)$ are the Floquet eigen energies and Floquet eigen states. Floquet theorem asserts that $\phi(\bf{r},t)$ is also a periodic function which has same periodicity as time dependent potential {\it i.e} $\phi(\bf{r},t)=\phi(\bf{r},t+T)$ with period $T=2\pi/\omega$. 
A unitary scattering matrix can be formed between the incident waves with energy $E$ and scattered waves with energies $ E+n\hbar \omega $ (given in \ref{FSMM}). 

We define  $\varepsilon=E/E_{0}$ and $v_{g}=V_{0}/E_{0}$, $v_{g}$ being  the effective barrier strength of the time independent potential  and all other energies are measured in terms of $ E_{0} =\hbar v_{F}/d$.  We measure $\omega$ in terms of $ v_F/d $. The large bulk gap of TI’s restrict the frequency and the amplitude range of the THz field. $ V_t $, $ V_0 $ and $ \hbar \omega $ should be less than the bulk band gap ($ V_{G}^{bulk} $) of the 3DTI. $ V_{G}^{bulk} $ is of the order of 300 meV for $Bi_{2}Se_{3}$ and 150 meV for $Bi_{2}Te_{3}$. We consider the low energy Dirac model of the surface states (described by Eq. \ref{eq:sch-equ}) where the contribution of the bulk electronic states can be omitted. In this work, we explore the drive strength (Z) regime where $ Z $ varies from 0.25 to 8.  We obtain interesting results by considering fermi energy close to the Dirac point which can be done by tuning the back gate. In this case, energy of a quantum of photon ($ \hbar \omega $) and the fermi energy are of the same order.

We obtain the transmission coefficient ($t_{n}$) by solving the Floquet scattering matrix (given in \ref{FSMM}) numerically. The total transmission probability 
\beq
T=\sum_{n}\frac{ cos(\theta_{n,R})}{cos(\theta_{I})}\vert t_{n}\vert^{2}
\label{eq:transmission}
\eeq
where the transmission probability $T_{n}=\frac{ cos(\theta_{n,R})}{cos(\theta_{I})}\vert t_{n}\vert^{2}$ defines the probability that an electron injected at the left electrode with energy $  E$ will transfer to the right electrode with energy $E+n\hbar\omega$. Eq. \ref{eq:transmission} contains infinite number of sum although we can truncate the infinite sum to the finite sum depending  on the value of $Z=\frac{V_{t}}{\hbar\omega}$. Z determines the coupling streangth of the time dependent potential with the system through Bessel function $J_{n}(V_{t}/\hbar \omega)$. We truncated the sum upto $ [-N,N]  $ where $\vert N \vert >\frac{V_{t}}{\hbar\omega}$.   
The zero temperature conductance is given by Landauer-B$\ddot{u}$ttiker formalism \cite{Landauer-a, Datta}
\beq
G=G_{0} \int_{-\pi/2}^{\pi/2} d\theta cos\theta ~T(\theta,E_{F})
\label{eq:cond}
\eeq
where $G_{0}=\frac{e^{2}k_{F}L_{y}}{\pi h}$. 
\section{Conclusions}\label{conclu}
To conclude, we have devised a scheme of using an incident THz beam to modulate the conductance of the surface states of a 3D topological insulator material. We have shown that this scheme can be used to build a photodetector (the domain of $V_t$ is 0.5 meV to 8 meV ) and that typical THz radiation would have sufficient drive strengths (Z varies from 0.25 to 8 ) to cause an observable modulation in observed conductance. The photodetector could be operated either by varying the bias voltages (bias voltage range is 0 to 10 meV ) applied to the device to do surface state G-V measurements or be operated in a scanning frequency mode to do spectrum measurements over a wide range of THz frequencies (explored THz range is 1 to 10 THz). The system could also be used as a sensitive probe for comparing elastic and inelastic scattering processes inside the 3DTI material. 

We observe the following appearances.  We show that the conduction gets modified in the presence of a THz radiation and takes distorted Breit-Wigner form as the drive strength becomes stronger. This is due to increase of the Floquet sidebands scattering processes resulting in the appearance of extra resonance peaks. We demonstrate that by increasing the bias between left and right contacts, it is possible to reduce the central band scattering process leaving behind only the Floquet sideband scattering processes. We observe that the conductance as a function of frequency shows the discontinuity at strong drive strength when the number of propagating modes in a specific Floquet band vanishes. For weak drive strength, the conductance does not show any discontinuity. The appearance of the discontinuity in the conductance can be used as a finite frequency noise detector.

P. Mondal  is supported by a UGC fellowship. 
\section{Appendix}
\subsection{Solution of the time dependent Dirac equation}
\subsubsection{Inside the barrier:}
The time dependent Dirac equation in presence of potential profile (Eq. \ref{pot}) becomes
\bea
i\hbar \frac{\partial \Psi(\bf{r},t)}{\partial t} &=& \hbar v_{F}\left(\begin{array}{cc}
 0 & -i\partial_{y}+\partial_{x}\\
  -i\partial_{y}-\partial_{x} & 0
\end{array} \right)\Psi(\bf{r},t)\nn\\
&+& V(x,t)\Psi(\bf{r},t)
\label{sch-equ}
\eea 
Eq. \ref{sch-equ} is separable inside the barrier due to space homogeneity of the potential V(x,t). So we can write the solution of Eq. \ref{sch-equ} as $\phi(\bf{r},t)=\psi(\bf{r}) f(t)$. $\psi(\bf{r})$ and $f(t)$ are the solutions of space and time dependent Dirac equation
\bea
 &&\hbar v_{F}\left(\begin{array}{cc}
 0 & -i\partial_{y}+\partial_{x}\\
  -i\partial_{y}-\partial_{x} & 0
\end{array} \right)\psi(\bf{r})+V_{0}\psi(\bf{r})= E \psi(\bf{r})\nn\\
\label{space-part}  \\
&& i\hbar \partial_{t} f(t)-V_{t} cos(\omega t)f(t)=(E-E_{F}) f(t)
 \label{time-part}
\eea
From Eq. \ref{time-part}, we obtain solution of the time dependent part as 
\bea
f(t) &=& e^{-i(E-E_{F})t/\hbar}e^{-i/\hbar \int_{0}^{t}V_{t}cos(\omega t^{'})} dt^{'}\nonumber\\
 &=& e^{-i(E-E_{F})t/\hbar} \sum_{n=-\infty}^{n=\infty}J_{n}(\frac{V_{t}}{\hbar\omega})e^{-in\omega t}
 \label{time-part1}  
\eea
where $J_{n}$ is the $n^{th}$ order Bessel function and is a function of $V_t$ and $\omega$. The strength of interaction  between the time dependent potential and system is defined by the dimensionless quantity  $Z=\frac{V_{t}}{\hbar \omega}$ . Since f(t) is a periodic function {\it i.e} f(t)=f(t+T), one can obtain from Eq. \ref{time-part1} that solution remains unchanged by changing $E=E_{F}+m\hbar\omega$, m is an integer.\\
We write Eq. \ref{space-part} in a dimensionless form as  

\bea
&& \left(\begin{array}{cc}
 0 & -i\bar{\partial}_{y}+\bar{\partial}_{x}\\
  -i\bar{\partial}_{y}-\bar{\partial}_{x} & 0
\end{array} \right)\psi(\bar{\bf{r}})=(\varepsilon -v_{g}) \psi(\bar{\bf{r}}) 
\eea
where $\hbar v_{F}/d =E_{0}  $ is the unit of the energy,  $\varepsilon=E/E_{0}$ and $v_{g}=V_{0}/E_{0}$. $v_{g}$ describe the effective barrier strength of potential barrier. 

\bea
(-i\bar{\partial}_{y}+\bar{\partial}_{x}) \psi_2(\bar{x},\bar{y}) &=&(\varepsilon -v_{g}) \psi_1(\bar{x},\bar{y})\nn\\
(-i\bar{\partial}_{y}-\bar{\partial}_{x}) \psi_1(\bar{x},\bar{y}) &=&(\varepsilon -v_{g}) \psi_2(\bar{x},\bar{y})\nn\\
\eea

The system is translationally invariant along the y direction, hence  $\psi_(\bar{x},\bar{y})=\phi(\bar{x})e^{iq_y \bar{y}}$ 
\bea
\bigg [\frac{\partial^{2}}{\partial \bar{x}^{2}}+\bigg({(\varepsilon_F-v_{g}+m\bar{E}_{t})^2-q_y^2\bigg)}\bigg ]\phi_2(\bar{x})=0\nn
\eea
where we have written  $\varepsilon=\varepsilon_F+m\bar{E}_{t}$ and  $\bar{E}_{t} = \hbar \omega/ E_0$. 
\bea
\phi_2(\bar{x})=\sum_m (A_m e^{ik_x^m \bar{x}}+B_m e^{-ik_x^m \bar{x}})\nn
\eea
where $(k_x^m)^2=(\varepsilon_F-v_{g}+m\bar{E}_{t})^2-q_y^2 $. 
\bea
\phi_1(\bar{x}) &=&\sum_m \frac{(q_y+ik_x^m)}{(\varepsilon_F-v_{g}+m\bar{E}_{t})}A_m e^{ik_x^m \bar{x}}\nn\\
&+&\frac{(q_y-ik_x^m)}{(\varepsilon_F-v_{g}+m\bar{E}_{t})}B_m e^{-ik_x^m \bar{x}}\nn\\
&=&  \sum_m i e^{-i\theta_m } A_m e^{ik_x^m \bar{x}} + (- i) e^{i\theta_m } B_m e^{-ik_x^m \bar{x}} \nn
\eea
and  $\theta_m=sin^{-1} (q_y/(\varepsilon_F-v_{g}+m\bar{E}_{t}))$ and  $k_x^m=(\varepsilon_F-v_{g}+m\bar{E}_{t}) cos (\theta_m)$. The solution remains unchanged if $\varepsilon_F$ shifted by $m\hbar\omega$. So, we choose $\varepsilon_F=\varepsilon$ .

We obtain solution of the time independent part of the Dirac equation as
\bea
\psi(\bar{\bf{r}})&=&\frac{1}{\sqrt{2}}\left[\sum_{m} A_{m} \left(\begin{array}{cc}
   e^{i(\pi/2-\theta_{m})}\\
  1 
\end{array} \right)e^{i[k_{x}^{m}\bar{x}+q_{y}\bar{y}]}\right]  \nonumber \\
&+& \frac{1}{\sqrt{2}}\left[\sum_{m} B_{m} \left(\begin{array}{cc}
  e^{-i(\pi/2-\theta_{m})}\\
  1 
\end{array} \right)e^{i[-k_{x}^{m}\bar{x}+q_{y}\bar{y}]} \right] \nn
\label{space-waveI}\\
\eea

Therefore, the full wavefunction (Eq. \ref{floquet}) inside the barrier is of the form   
\bea
\Psi_{II_a}(\bar{\bf{r}},\bar{t})&=&\frac{e^{-i\varepsilon \bar{t} }}{\sqrt{2}}\sum_{n,m} A_{m} \left(\begin{array}{cc}
  e^{i(\pi/2-\theta_{m}^{1})}\\
  1 
\end{array} \right)e^{i[k_{x}^{m}\bar{x}+q_{y}\bar{y}]} \nonumber \\
 & \times & e^{-i(m+n)\bar{\omega} \bar{t}}J_{m}(\frac{\bar{V}_{t}}{\bar{E}_t})  \nonumber \\
&+& \frac{e^{-i \varepsilon \bar{t} }}{\sqrt{2}}\sum_{n,m} B_{m} \left(\begin{array}{cc}
  e^{-i(\pi/2-\theta_{m}^{1})}\\
  1 
\end{array} \right)e^{i[-k_{x}^{m}\bar{x}+q_{y}\bar{y}]}\nn\\
& \times & e^{-i(m+n)\bar{\omega} \bar{t}}J_{m}(\frac{\bar{V}_{t}}{\bar{E}_t})
\label{waveIIa}
\eea
where we define $ \bar{t}=t v_{F}/d $, $\bar{\omega} =  \omega d/v_{F}$, $ \bar{V}_{t}=V_t/E_0  $,  $\theta_{m}^{1}=sin^{-1}\bigg( \frac{q_{y}}{\varepsilon-v_{g}+ m\bar{E}_{t} }\bigg)$ and $k_{x}^{m}=(\varepsilon-v_{g}+m\bar{E}_{t}) cos(\theta_{m}^{1})$.
\bea
\Psi_{II_b}(\bar{\bf{r}},\bar{t})&=&\frac{e^{-i\varepsilon \bar{t} }}{\sqrt{2}}\sum_{n,m} C_{m} \left(\begin{array}{cc}
 e^{i(\pi/2-\theta_{m}^{2})}\\
  1 
\end{array} \right)e^{i[q_{x}^{m}\bar{x}+q_{y}\bar{y}]}\nn\\
 &\times & e^{-i(m+n)\bar{\omega} \bar{t}}J_{m}(\frac{\bar{V}_{t}}{\bar{E}_t})  \nonumber \\
&+& \frac{e^{-i \varepsilon \bar{t}}}{\sqrt{2}}\sum_{n,m} D_{m} \left(\begin{array}{cc}
 e^{-i(\pi/2-\theta_{m}^{2})}\\
  1 
\end{array} \right)e^{i[-q_{x}^{m}\bar{x}+q_{y}\bar{y}]} \nn\\
 &\times & e^{-i(m+n)\bar{\omega} \bar{t}}J_{m}(\frac{\bar{V}_{t}}{\bar{E}_t})
\label{waveIIb}
\eea
where  $\theta_{m}^{2}=sin^{-1}\bigg( \frac{q_{y}}{ \varepsilon + m\bar{E}_t}\bigg)$, and $q_{x}^{m}=(\varepsilon +m\bar{E}_t)cos(\theta_{m}^{2})$.
\subsubsection{Outside the barrier:}
Electrons outside the barrier region absorb /emit $  n$ quanta from/to the incident radiation when passing through or reflected by the potential. So the energy E of the incident electrons change to the energy $E+n\hbar\omega$ of the reflected/transmitted electrons where n=$0,\pm 1,\pm 2$. So, the incident and reflected wavefunction can be written as
\bea
\Psi_{I}(\bar{\bf{r}},\bar{t}) &=&\Psi_{i}(\bar{\bf{r}},\bar{t})+\Psi_{r}(\bar{\bf{r}},\bar{t}) \nonumber\\
&=&\frac{e^{-i\varepsilon \bar{t}}}{\sqrt{2}}\left[\left(\begin{array}{cc}
  e^{i(\pi/2-\theta_{0}^{L})}\\
  1 
\end{array} \right)e^{i[q_{x,L}\bar{x}+q_{y}\bar{y}]} \right]\nonumber \\
&+&\frac{1}{\sqrt{2}}\sum_{n} r_{n} \left(\begin{array}{cc}
  e^{-i(\pi/2-\theta_{n}^{L})}\\
  1 
\end{array} \right)e^{i[-q_{x,L}^{n}\bar{x}+q_{y}\bar{y}]}\nn\\
&\times & e^{-i(\varepsilon+ n \bar{E}_{t}) \bar{t}}
\eea 
where $\bar{\mu}_{L}=\mu_{L} /E_0$, $q_{x,L}=( \varepsilon-\bar{\mu}_{L} ) cos(\theta_{0}^{L})$, $q_{y}=( \varepsilon-\bar{\mu}_{L}) sin(\theta_{0}^{L})$ and $\theta_{0}^{L}=sin^{-1}(q_{y}/( \varepsilon-\bar{\mu}_{L}))$. $\theta_{n}^{L}=sin^{-1}(q_{y}/(\varepsilon-\bar{\mu}_{L} + n \bar{E}_{t}))$,  and $q_{x,L}^{n}=( \varepsilon-\bar{\mu}_{L}+n\bar{E}_{t})cos(\theta_{n}^{L})$.
 
The transmitted wavefunction becomes 
\bea
\Psi_{t}(\bar{\bf{r}},\bar{t}) &=& \frac{e^{-i\varepsilon \bar{t}}}{\sqrt{2}}\sum_{n} t_{n} \left(\begin{array}{cc}
e^{i(\pi/2-\theta_{n}^{R})}\\
  1 
\end{array} \right)e^{i[q_{x,R}^{n}\bar{x}+q_{y}\bar{y}]} \nn\\
& \times & e^{-i(\varepsilon+ n\bar{E}_{t})\bar{t}} 
\eea
where we define $\theta_{n}^{R}=sin^{-1}(q_{y}/(\varepsilon-\bar{\mu}_{R} + n \bar{E}_{t}))$,  and $q_{x,R}^{n}=( \varepsilon-\bar{\mu}_{R}+n\bar{E}_{t}) cos(\theta_{n}^{R})$.
\begin{figure} [h!]
\includegraphics[width=1\columnwidth,height=0.55 
\columnwidth]{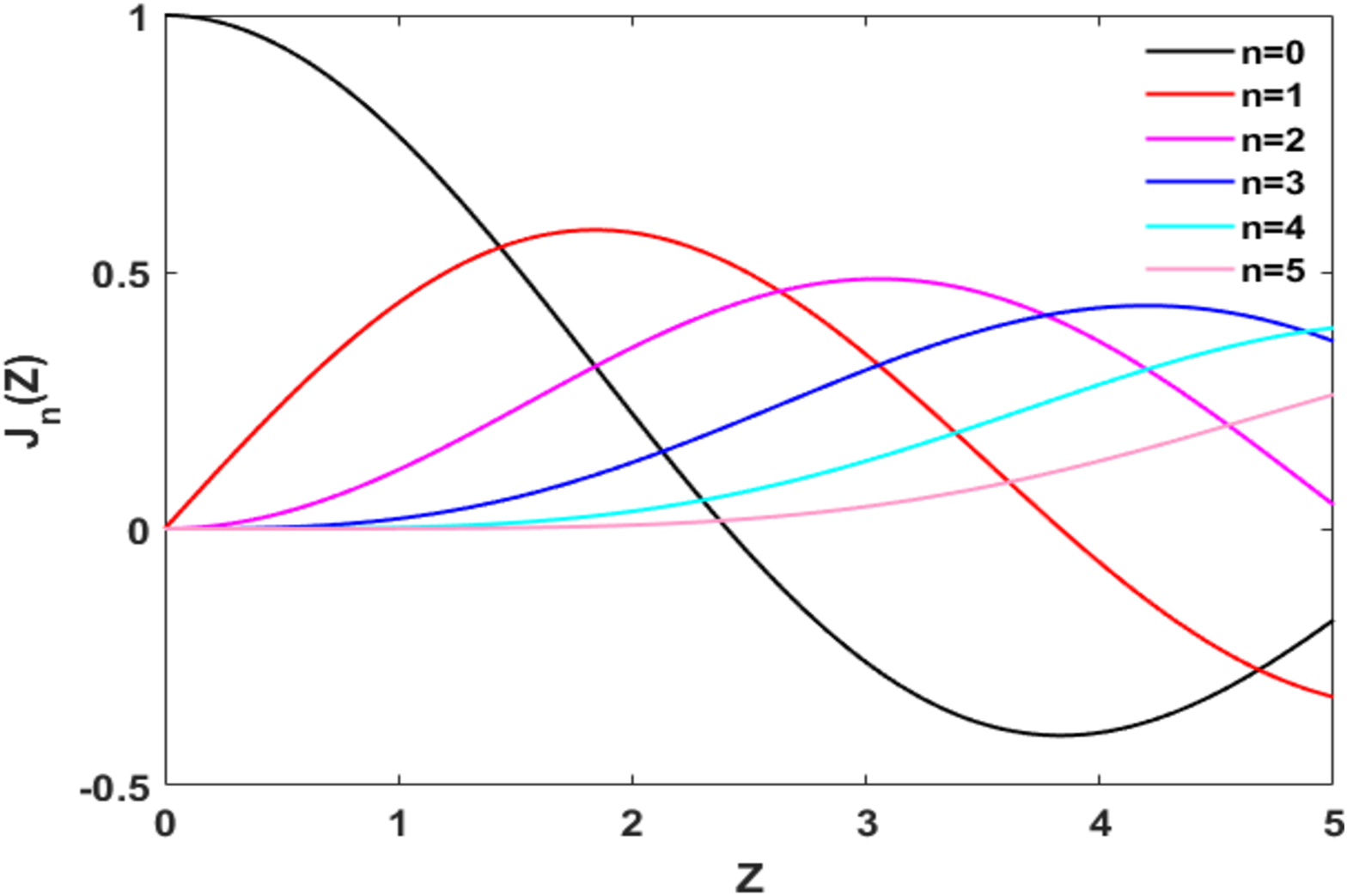} 
\caption{Bessel function variation with respect to Z and n.}
\label{fig:bessel}
\end{figure}
\subsubsection{Zero energy states solution} \label{zero_energy}
The solution of the wavefunction  outside the barrier when $ \varepsilon-\bar{\mu}_{L}(\bar{\mu}_{R}) + n \bar{E}_{t}=0 $
\bea
(-i\bar{\partial}_{y}+\bar{\partial}_{x}) \psi_2(\bar{x},\bar{y}) &=&0 \nn\\
(-i\bar{\partial}_{y}-\bar{\partial}_{x}) \psi_1(\bar{x},\bar{y}) &=&0 \nn
\eea
\bea
\phi_1(\bar{x}) &= & a_{1} e^{ q_y \bar{x}}\nn\\
\phi_2(\bar{x}) &= & a _{2}e^{- q_y \bar{x}}
\label{eq:boundstates_sol}
\eea
Therefore, Eq. \ref{eq:boundstates_sol} is the solution of the bound states which gives zero transmission/reflection coefficient. The corresponding transmission probability is discussed in the main text (Eq. \ref{eq:zero_energy}). 
\subsection{Floquet scattering matrix method} \label{FSMM}
\subsubsection{Single barrier}
\begin{widetext}
Applying the Floquet matrix method \cite{Li} at the boundaries

At $ \bar{x}=-1/2 $

\bea
 e^{i(\pi/2-\theta_{n}^{L})}e^{-iq_{x,L}^{n}(1/2)}I_n+  e^{-i(\pi/2-\theta_{n}^{L})}e^{iq_{x,L}^{n}(1/2)}r_n &=& \sum_m J_{n-m}(\alpha)[ e^{i(\pi/2-\theta_{m}^{1})}e^{-ik_{x}^{m}(1/2)} A_m+ e^{-i(\pi/2-\theta_{m}^{1})}e^{ik_{x}^{m}(1/2)} B_m]\nn\\
e^{-iq_{x,L}^{n}(1/2)} I_n+ e^{iq_{x,L}^{n}(1/2) } r_n &=&  \sum_mJ_{n-m}(\alpha)[ e^{-ik_{x}^{m}(1/2)} A_m + e^{ik_{x}^{m}(1/2)} B_m]\nn
\eea
where $ I_{n}=\delta_{n,0} $ and the equation can be written as 
\bea
\sum_m[ e^{i(\pi/2-\theta_{m}^{L})}e^{-iq_{x,L}^{m}(1/2)}I_m + e^{-i(\pi/2-\theta_{m}^{L})}e^{iq_{x,L}^{m}(1/2)}r_m]\delta_{m,n} &=& \sum_m J_{n-m}(\alpha)[ e^{i(\pi/2-\theta_{m}^{1})}e^{-ik_{x}^{m}/2} A_m+ e^{-i(\pi/2-\theta_{m}^{1})}e^{ik_{x}^{m}/2} B_m]\nn\\
\sum_m[e^{-iq_{x,L}^{m}}(1/2) I_m+ e^{iq_{x,L}^{m}(1/2) } r_m]\delta_{m,n} &=&  \sum_mJ_{n-m}(\alpha)[ e^{-ik_{x}^{m}/2} A_m + e^{ik_{x}^{m}/2} B_m]\nn
\eea
We can write the above equations in a matrix form as follows
\bea
&&\left(\begin{array}{cc}
 e^{i(\pi/2-\theta_{m}^{L})}e^{-iq_{x,L}^{m}(1/2)}\delta_{m,n} & ~~ e^{-i(\pi/2-\theta_{m}^{L})}e^{iq_{x,L}^{m}(1/2)}\delta_{m,n}\\
e^{-iq_{x,L}^{m}(1/2)}\delta_{m,n} & e^{iq_{x,L}^{m}(1/2) }\delta_{m,n}   
\end{array} \right)\left(\begin{array}{cc}
I_{m}\\
r_{m}
\end{array}\right)\nonumber\nn\\
&=& \left(\begin{array}{cc}
 e^{i(\pi/2-\theta_{m}^{1})}e^{-ik_{x}^{m}(1/2)}J_{n-m}(\alpha) & ~~e^{-i(\pi/2-\theta_{m}^{1})}e^{ik_{x}^{m}(1/2)}J_{n-m}(\alpha)\\
e^{-ik_{x}^{m}(1/2)}J_{n-m}(\alpha) & e^{ik_{x}^{m}(1/2)}J_{n-m}(\alpha)(\alpha)   
\end{array} \right)\nonumber\\
&\times & \left(\begin{array}{cc}
A_{m}\\
B_{m}
\end{array}\right)
\label{eq:fq1}
\eea
where each component in the matrix is a $ n \times m $ order matrix. 
We consider $M_{\theta_{L(R)}}= \left(\begin{array}{cc}
e^{i(\pi/2-\theta_{m}^{L(R)})}\delta_{m,n} & e^{-i(\pi/2-\theta_{m}^{L(R)})}\delta_{m,n}\\
\delta_{m,n} & \delta_{m,n} 
\end{array}\right)$,
 $M_{L(R)}= \left(\begin{array}{cc}
e^{-iq_{x,L(R)}^{m}(1/2)}\delta_{m,n} & 0\\
0 & e^{iq_{x,L(R)}^{m}(1/2)}\delta_{m,n} 
\end{array}\right)$,
$M_{\theta_{1}}= \left(\begin{array}{cc}
e^{i(\pi/2-\theta_{m}^{1})}\delta_{m,n} & e^{-i(\pi/2-\theta_{m}^{1})}\delta_{m,n}\\
\delta_{m,n} & \delta_{m,n}
\end{array}\right)$, $M_{B}= \left(\begin{array}{cc}
e^{-ik_{x}^{m}(1/2)}\delta_{m,n}& 0\\
0 & e^{ik_{x}^{m}(1/2)}\delta_{m,n} 
\end{array}\right)$, $M_{J}= \left(\begin{array}{cc}
J_{n-m}(\alpha) & 0\\
0 & J_{n-m}(\alpha)
\end{array}\right)$. Here, $M_{AL(R)}^{\ast}=M_{AL(R)}^{-1}$ and $M_{B}^{\ast}=M_{B}^{-1}$

Eq. \ref{eq:fq1} can be written in terms of the above defined variables
\bea
 (M_{\theta_{L}} M_{L}) \left(\begin{array}{cc}
I_{m}\\
r_{m}
\end{array}\right)=  (M_{J} M_{\theta_{1}} M_{B})\left(\begin{array}{cc}
A_{m}\\
B_{m}
\end{array}\right)
\eea

At $ \bar{x}=1/2 $
\bea
 \sum_m J_{n-m}(\alpha)[ e^{i(\pi/2-\theta_{m}^{1})}e^{ik_{x}^{m}(1/2)} A_m+ e^{-i(\pi/2-\theta_{m}^{1})}e^{-ik_{x}^{m}(1/2)} B_m] &=&   e^{i(\pi/2-\theta_{n}^{R})}e^{iq_{x,R}^{n}(1/2)} t_n  \nn\\
 \sum_mJ_{n-m}(\alpha)[ e^{ik_{x}^{m}(1/2)} A_m + e^{-ik_{x}^{m}(1/2)} B_m] &=&  e^{iq_{x,R}^{n}(1/2)}t_n
\eea
The above equation can be written as
\bea
 \sum_m J_{n-m}(\alpha)[ e^{i(\pi/2-\theta_{m}^{1})}e^{ik_{x}^{m}(1/2)} A_m+ e^{-i(\pi/2-\theta_{m}^{1})}e^{-ik_{x}^{m}(1/2)} B_m] &=& \sum_m   e^{i(\pi/2-\theta_{m}^{R})}e^{iq_{x,R}^{m}(1/2)} t_m \delta_{m,n}   \nn\\
 \sum_m J_{n-m}(\alpha)[ e^{ik_{x}^{m}(1/2)} A_m + e^{-ik_{x}^{m}(1/2)} B_m] &=& \sum_m  e^{iq_{x,R}^{m}(1/2)}t_m \delta_{m,n} 
 \label{eq:fq2} 
\eea
We can write Eq. \ref{eq:fq2} as 
\bea
&& \left(\begin{array}{cc}
 e^{i(\pi/2-\theta_{m}^{1})}e^{ik_{x}^{m}(1/2)}J_{n-m}(\alpha) & ~~ e^{-i(\pi/2-\theta_{m}^{1})}e^{-ik_{x}^{m}(1/2)}J_{n-m}(\alpha)\\
e^{ik_{x}^{m}(1/2)}J_{n-m}(\alpha) & e^{-ik_{x}^{m}(1/2)}J_{n-m}(\alpha) 
\end{array} \right)\nn\\
&\times & \left(\begin{array}{cc}
A_{m}\\
B_{m}
\end{array} \right)\nn\\
&=&  \left(\begin{array}{cc}
 e^{i(\pi/2-\theta_{m}^{R})}e^{iq_{x,R}^{m}(1/2)}\delta_{m,n} & ~~ e^{-i(\pi/2-\theta_{m}^{R})}e^{-iq_{x,R}^{m}(1/2)}\delta_{m,n}\\
e^{iq_{x,R}^{m}(1/2)}\delta_{m,n} & e^{-iq_{x,R}^{m}(1/2) }\delta_{m,n}   
\end{array} \right)
\left(\begin{array}{cc}
t_{m}\\
0
\end{array}\right)
\label{eq:fq3}
\eea
We can write Eq. \ref{eq:fq3} using the defined variables as 
\bea
 (M_{J}M_{\theta_{1}} M_{B}^{\ast})\left(\begin{array}{cc}
A_{m}\\
B_{m}
\end{array}\right) =  (M_{\theta_{R}} M_{R}^{\ast}) \left(\begin{array}{cc}
t_{m}\\
0
\end{array}\right) 
\eea

So equations take form 
\bea
 (M_{\theta_{L}} M_{L}) \left(\begin{array}{cc}
I_{m}\\
r_{m}
\end{array}\right)&=&  (M_{J}M_{\theta_{1}} M_{B})\left(\begin{array}{cc}
A_{m}\\
B_{m}
\end{array}\right)\nn\\
 (M_{J} M_{\theta_{1}} M_{B}^{\ast})\left(\begin{array}{cc}
A_{m}\\
B_{m}
\end{array}\right) &=&  (M_{\theta_{R}} M_{R}^{\ast}) \left(\begin{array}{cc}
t_{m}\\
0
\end{array}\right)
\eea

Thus, the Floquet transfer matrix is of the form
\bea
M_{TF}&=& M_{L}^{-1} M_{\theta_{L}}^{-1}(M_{J}M_{\theta_{1}} M_{B})( M_{B} M_{\theta_{1}}^{-1} M_{J}^{-1} )(M_{\theta_{R}} M_{R}^{\ast})\nn\\
&=& (M_{L}^{\ast} M_{\theta_{L}}^{-1}) (M_{J}M_{\theta_{1}} M_{B}^{2} M_{\theta_{1}}^{-1} M_{J}^{-1})(M_{\theta_{R}} M_{R}^{\ast})
\eea
\subsubsection{Double barrier potential}
We consider boundaries at $ \bar{x}=-1-b/2d,-b/2d,b/2d,1+b/2d  $. In a similar way, we obtain Floquet transfer matrix for double barrier potential.
At the boundary $ \bar{x}=-1-b/2d $, 

\bea
\left(\begin{array}{cc}
 e^{i(\pi/2-\theta_{n}^{L})}e^{-iq_{x,L}^{n}(1+b/2d)} \delta_{m,n} & ~~  e^{-i(\pi/2-\theta_{n}^{L})}e^{iq_{x,L}^{n}(1+b/2d)} \delta_{m,n}\\
e^{-iq_{x,L}^{n}(1+b/2d)} \delta_{m,n} & e^{iq_{x,L}^{n}(1+b/2d) \delta_{m,n}}   
\end{array} \right)\left(\begin{array}{cc}
I_{n}\\
r_{n}
\end{array}\right)\nonumber
\eea
\bea
&=& \left(\begin{array}{cc}
 e^{i(\pi/2-\theta_{m}^{1})}e^{-ik_{x}^{m}(1+b/2d)}J_{n-m}(\alpha) & e^{-i(\pi/2-\theta_{m}^{1})}e^{ik_{x}^{m}(1+b/2d)}J_{n-m}(\alpha)\\
e^{-ik_{x}^{m}(1+b/2d)}J_{n-m}(\alpha) & e^{ik_{x}^{m}(1+b/2d)}J_{n-m}(\alpha)
\end{array} \right)\nonumber\\
&\times &\left(\begin{array}{cc}
A_{m}\\
B_{m}
\end{array}\right)\nonumber
\eea
Boundary condition at $ \bar{x}=-b/2d $ 
\bea
& &\left(\begin{array}{cc}
 e^{i(\pi/2-\theta_{m}^{1})}e^{-ik_{x}^{m}(b/2d)}J_{n-m}(\alpha) & ~~ e^{-i(\pi/2-\theta_{m}^{1})}e^{ik_{x}^{m}(b/2d)}J_{n-m}(\alpha)\\
e^{-ik_{x}^{m}(b/2d)}J_{n-m}(\alpha) & e^{ik_{x}^{m}(b/2d)}J_{n-m}(\alpha)(\alpha)   
\end{array} \right)\nonumber\\
&\times& \left(\begin{array}{cc}
A_{m}\\
B_{m}
\end{array}\right)\nonumber
\eea
\bea
&=  &\left(\begin{array}{cc}
 e^{i(\pi/2-\theta_{m}^{2})}e^{-iq_{x}^{m}(b/2d)}J_{n-m}(\alpha) & ~~ e^{-i(\pi/2-\theta_{m}^{2})}e^{iq_{x}^{m}(b/2d)}J_{n-m}(\alpha)\\
e^{-iq_{x}^{m}(b/2d)}J_{n-m}(\alpha) & e^{iq_{x}^{m}(b/2d)}J_{n-m}(\alpha)(\alpha)   
\end{array} \right)\nonumber\\
&\times & \left(\begin{array}{cc}
C_{m}\\
D_{m}
\end{array}\right)\nonumber
\eea
Boundary condition at $ \bar{x}=b/2d  $
\bea
& &\left(\begin{array}{cc}
 e^{i(\pi/2-\theta_{m}^{2})}e^{iq_{x}^{m}(b/2d)}J_{n-m}(\alpha) & ~~ e^{-i(\pi/2-\theta_{m}^{2})}e^{-iq_{x}^{m}(b/2d)}J_{n-m}(\alpha)\\
e^{iq_{x}^{m}(b/2d)}J_{n-m}(\alpha) & e^{-iq_{x}^{m}(b/2d)}J_{n-m}(\alpha)(\alpha)   
\end{array} \right)\nonumber\\
&\times & \left(\begin{array}{cc}
C_{m}\\
D_{m}
\end{array}\right)\nonumber
\eea
\bea
& =  &\left(\begin{array}{cc}
 e^{i(\pi/2-\theta_{m}^{1})}e^{ik_{x}^{m}(b/2d)}J_{n-m}(\alpha) & ~~ e^{-i(\pi/2-\theta_{m}^{1})}e^{-ik_{x}^{m}(b/2d)}J_{n-m}(\alpha)\\
e^{ik_{x}^{m}(b/2d)}J_{n-m}(\alpha) & e^{-ik_{x}^{m}(b/2d)}J_{n-m}(\alpha)(\alpha)   
\end{array} \right)\nonumber\\
&\times& \left(\begin{array}{cc}
E_{m}\\
F_{m}
\end{array}\right)\nonumber
\eea
Boundary condition at $ \bar{x}=1+b/2d $
\bea
&  &\left(\begin{array}{cc}
 e^{i(\pi/2-\theta_{m}^{1})}e^{ik_{x}^{m}(1+b/2d)}J_{n-m}(\alpha) & e^{-i(\pi/2-\theta_{m}^{1})}e^{-ik_{x}^{m}(1+b/2d)}J_{n-m}(\alpha)\\
e^{ik_{x}^{m}(1+b/2d)}J_{n-m}(\alpha) & e^{-ik_{x}^{m}(1+b/2d)}J_{n-m}(\alpha)(\alpha)   
\end{array} \right)\nonumber\\
&\times &\left(\begin{array}{cc}
E_{m}\\
F_{m}
\end{array}\right)\nonumber
\eea
\bea
&=&  \left(\begin{array}{cc}
 e^{i(\pi/2-\theta_{n}^{R})}e^{iq_{x,R}^{n}(1+b/2d)} & ~~  e^{-i(\pi/2-\theta_{n}^{L})}e^{-iq_{x,R}^{n}(1+b/2d)}\\
e^{iq_{x,R}^{n}(1+b/2d)} & e^{-iq_{x,R}^{n}(1+b/2d) }   
\end{array} \right)
\left(\begin{array}{cc}
t_{n}\\
0
\end{array}\right)
\label{tra-equ2}
\eea
The scattering matrix is given as

\bea
S_{11}(L,i+1)& =&[M_{11}-S_{11}(L,i) M_{21}(i, i+1)]^{-1} S_{11}(L,i)\nn\\
S_{12}(L,i+1)& =& [M_{11}-S_{12}(L, i) M_{21}(i,i+1)]^{-1}[S_{12}(L,i)M_{22}(i, i +1)-M_{12}(i, i+1)]\nn\\
S_{21}(L,i+1)& =& S_{22}(L, i)M_{21}(i, i +1)S_{11}(L, i +1)+S_{21}(L, i)\nn\\
S_{22}(L,i+1)& =& S_{22}(L, i)M_{21}(i, i+1)S_{12}(L,i+1)+ S_{22}(L, i )M_{22}( i, i +1 ) \nn
\eea
Where M's are the transfer matrices. The scattering matrix S(L,R)=S(L,N+1) is calculated by considering i =N and M(N, N+1)=M(N, R). The initial scattering matrix S(L, 1) can be obtained by taking i =0, M(0, 1)=M(L, 1) and S(L,0)=S(L,L) with S(L,L)=1. The transmission coefficient is given by $ t_{n}= S_{11}(L,R) I_{n}$. 
\end{widetext}

\subsection{Transmission of the Floquet side bands} 
\begin{figure*}[h!]
\subfloat{\includegraphics[width=2\columnwidth,height=0.60 
\columnwidth]{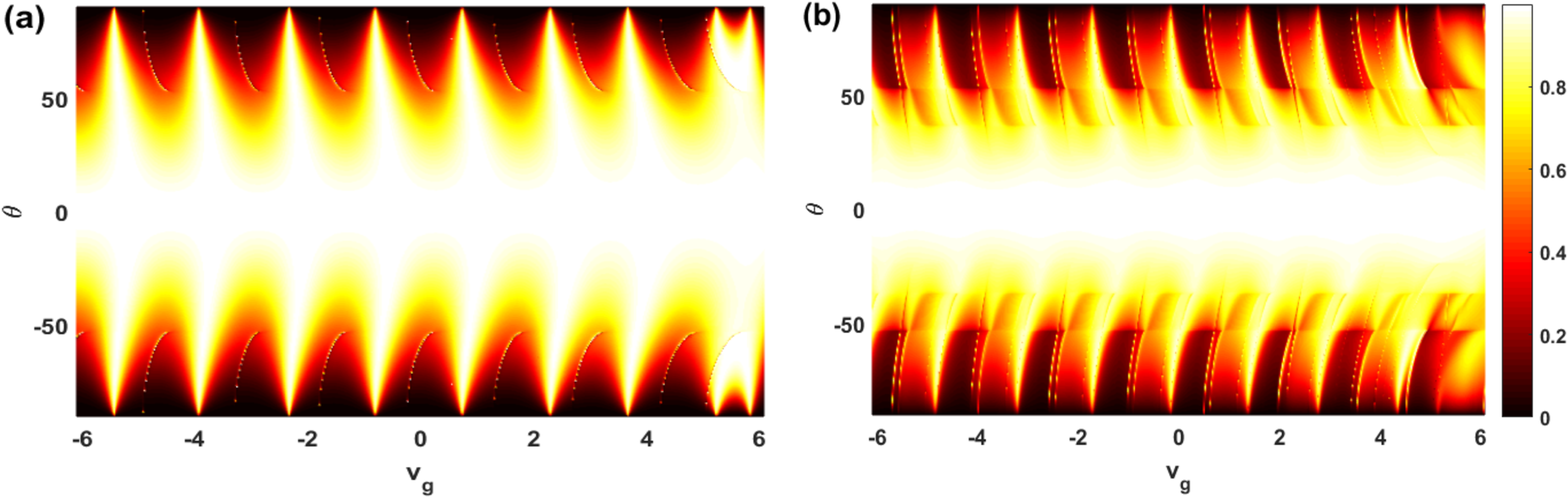} }
\caption{Transmission plot versus effective strength of scalar potential for a fixed frequency $\bar{\omega}=$0.305 and different values of amplitude of time dependent potential (a) $\bar{V}_{t}$=  0.076 (b) $\bar{V}_{t}$= 0.6088. The energy is measured in units of $ \hbar v_{F}/d $= 6.57 meV. Frequency is measured in units of $ v_{F}/d= 10^{13} $ Hz. We consider $\varepsilon=$ 7.610, $ b/d  $=2  and $ \bar{\mu}_{L}=\bar{\mu}_{R} $=6.088.}
\label{fig:tran1}
\end{figure*}

\begin{figure*}
\includegraphics[width=2\columnwidth,height=1
\columnwidth]{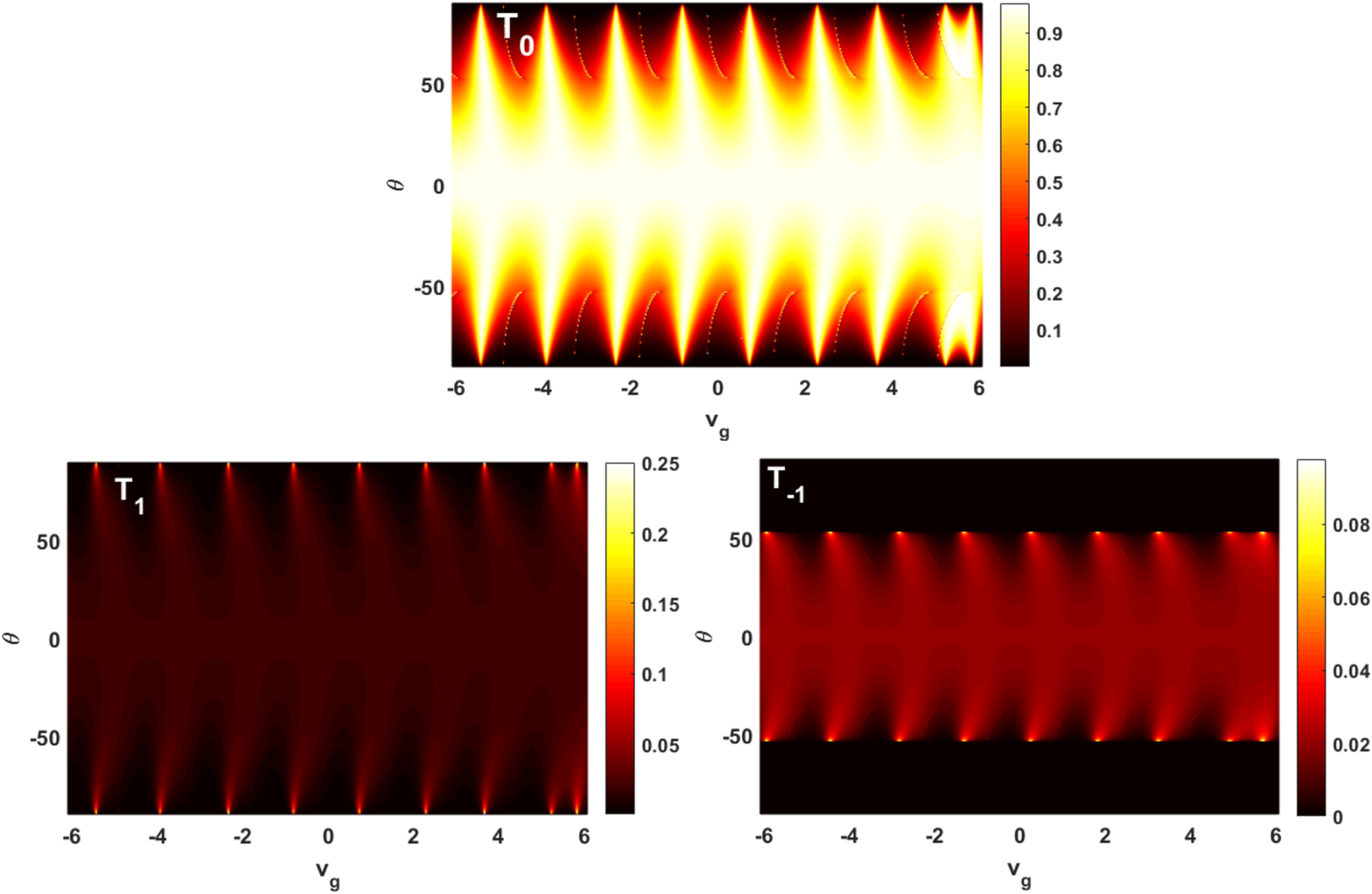} 
\caption{Transmission of the central band ($T_0$) and Floquet side bands ($T_{\pm 1}$) corresponding to Fig. \ref{fig:tran1} (a).}
\label{fig:tran_sideband_Vt05meV}
\end{figure*}

In Fig. \ref{fig:tran1}, we have shown total transmission vs angle of incidence and effective barrier strength for a fixed $Z=0.5$ (Fig. \ref{fig:tran1} a) and $Z=2$ (Fig. \ref{fig:tran1} b). In Fig. \ref{fig:tran1} a, the value of Z is 0.25 corresponding to the case of weaker drive strength. The total transmission follows the sum rule given in Eq. \ref{eq:transmission}. 
The transmission of side bands (inelastic scattering) and central band (elastic scattering) is plotted in the Fig. \ref{fig:tran_sideband_Vt05meV} for Fig. \ref{fig:tran1} (a). The condition of getting resonance condition in transmission is modified in a time dependent potential and given as $\varepsilon_{n}=\varepsilon_{r}+n \bar{E}_t  $ where $\varepsilon_{r}$ is the unperturbed resonance energy and $\bar{E}_t=\hbar \omega /v_g $.  The appearance of the extra peak in the $T_0$ can be understood from the fact that electron while passing through the system first emits a photon and then absorbs a photon such that the energy of the incident and transmitted electron remains same. But, this process involves absorption and emission of a photon which gives extra resonance peaks in $T_0$.

\begin{figure*}[h!]
\subfloat{\includegraphics[width=2\columnwidth,height=1.3
\columnwidth]{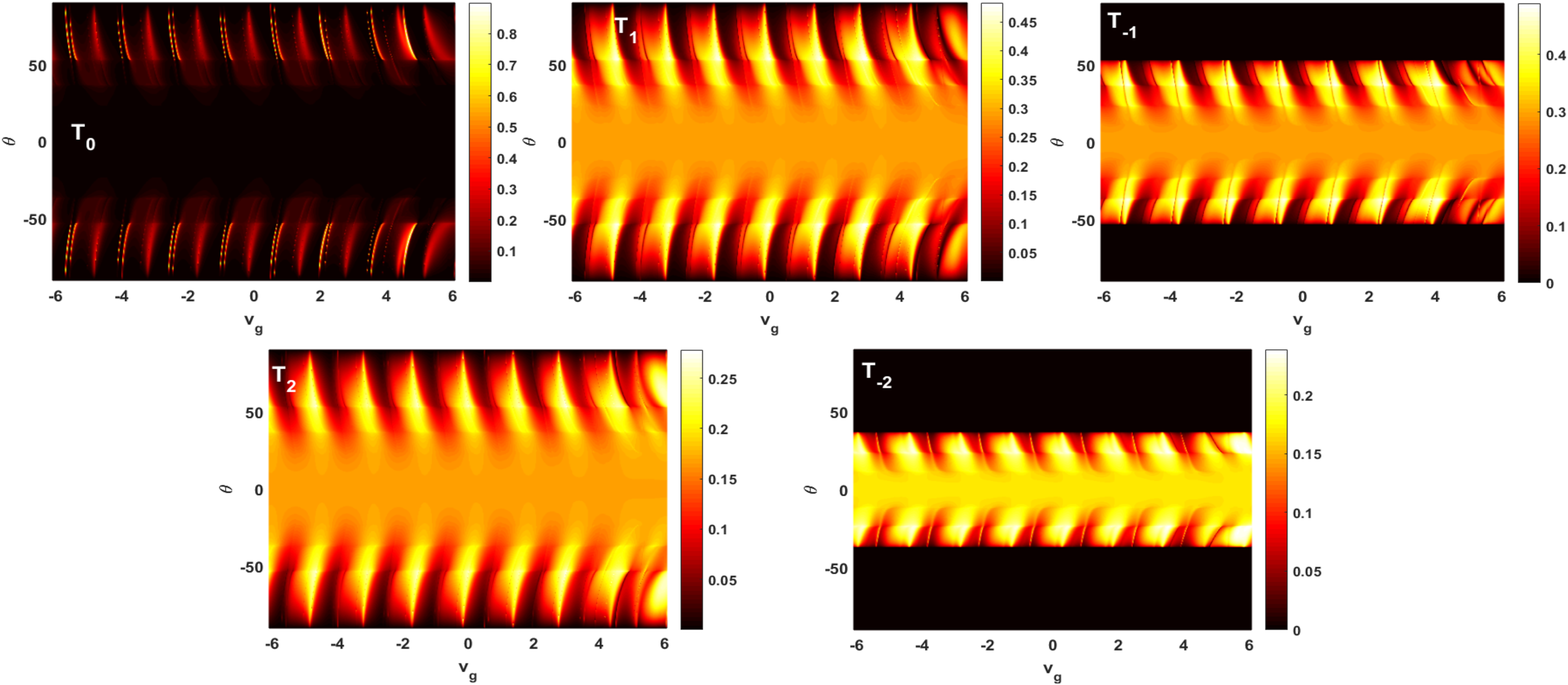} }
\caption{Transmission of the central band ($T_0$) and Floquet side bands ($T_{\pm 1, \pm 2}$) corresponding to Fig. \ref{fig:tran1} (b).}
\label{fig:tran_sideband_Vt4meV}
\end{figure*}

The contribution of the Floquet side bands to the transmission   ($T_{\pm 1}$) is small for Z=0.25 (shown in Fig. \ref{fig:tran_sideband_Vt05meV}).           
 Photon absorption process ($T_{1}$) is observed at both small and large angles of incidence where the electron energy changes from $\varepsilon$ to $\varepsilon+\bar{E}_t$ ($T_1$ in Fig. \ref{fig:tran_sideband_Vt05meV}) ). Whereas, the photon emission process ($T_{-1}$) is not observed at large angles of incidence ($T_{-1}$ in Fig. \ref{fig:tran_sideband_Vt05meV} and $T_{-1,-2}$ in Fig. \ref{fig:tran_sideband_Vt4meV}). The emission process is bounded by some critical angle ($\theta_{-n}^{R,c}$) above which the momentum of the transmitted electron ($q_{x,R}^{n}$) with energy $\varepsilon-\bar{\mu}_{R}-n\bar{E}_t$ becomes imaginary. This gives the evanescent solution of the electron. Thus, the evanescent solution occurs when $q_y > (\varepsilon-\bar{\mu}_{R} - n\bar{E}_{t}) $, from which one can calculate $\theta_{-n}^{R,c}$ as 
 \beq
 \theta_{-n}^{R,c}= arcsin\bigg((\varepsilon-\bar{\mu}_{R} -n\bar{E}_{t})/(\varepsilon-\bar{\mu}_{L}) \bigg) 
 \eeq
We calculate $\theta_{-n}^{R,c}$ = $53.13^{\circ}$ for $n=-1$ and $\theta_{-n}^{R,c}$ = $36.87^{\circ}$  for $n=-2$ Floquet bands. The angular spread of the transmission probability of the Floquet band $ T_{-n} $ gets reduced.   

The Floquet side band transmission corresponding to \ref{fig:tran1} (b) is plotted in the Fig. \ref{fig:tran_sideband_Vt4meV}. The Floquet side band scattering process increases as we increase Z while the central band scattering process gets reduced. 
At Z=2, $J_0(Z)$ is smaller than $|J_{\pm 1, 2}(Z)|$ (as shown in the Fig. \ref{fig:bessel}) which gives small contribution of the central band transmission to the total transmission.   


\newpage
\def \bb{\bibitem}

\end{document}